\documentstyle[twocolumn,aps,prl,array,epsf]{revtex}
\begin{document}
\draft
\title{Quantum key distribution in the Holevo limit}
\author{Ad\'{a}n Cabello\thanks{
Electronic address: adan@cica.es}}
\address{Departamento de F\'{\i}sica Aplicada II,
Universidad de Sevilla, 41012 Sevilla, Spain}

\date{\today}
\maketitle

\begin{abstract}
A theorem by Shannon and the Holevo theorem
impose
that the efficiency of any protocol for quantum key distribution,
$\cal E$, defined as the number of secret (i.e., allowing eavesdropping
detection) bits per transmitted bit plus qubit, is ${\cal E} \le 1$. The
problem addressed here is whether the limit ${\cal E} =1$ can be achieved.
It is showed that it can be done by splitting the secret bits
between several qubits and forcing Eve to have only a
sequential access to the qubits,
as proposed by Goldenberg
and Vaidman. A protocol
with ${\cal E} =1$ based on polarized photons and in which Bob's
state discrimination can be implemented with linear optical elements
is presented.
\end{abstract}



\narrowtext
In information theory one of the most fundamental questions is how
efficiently can one
transmit information by means of a given set of resources.
If this information is classical
(i.e., it can be expressed as a sequence of zeros and ones, or ``bits'')
a crucial theorem of classical information theory states that
if a (classical) communication channel
has mutual information $I(X\!:Y)$ between the input signal $X$ and
the received output $Y$, then that channel can be used to send up to,
but no more than,
$I(X\!:Y)$ bits \cite{Shannon49}.
The mutual information is defined as
\begin{equation}
I(X\!:Y)=H(X)-H (X | Y),
\end{equation}
where $H$ is the Shannon entropy, which is a function of the probabilities
$p(x_i)$ of the possible values of $X$, and is given by
$H(X)=-\sum\limits_i {p(x_i})\log _2p(x_i)$,
where the sum is over those $i$ with $p(x_i) > 0$.
$H (X | Y)$ is the expected entropy of $X$ once one
knows the value of $Y$, and is given by
\begin{equation}
H(X|Y)=\sum\limits_j {p(y_j)}\left[ {-\sum\limits_i {p(x_i|y_j})
\log _2p(x_i|y_j)}\right].
\end{equation}
A simple application of the above theorem reveals that using a
classical two-level system as a communication channel
(i.e., if the input signal $X$ can take only two values $x_0$ and $x_1$)
one is allowed to send up to, but no more than one bit [and
this occurs if $p(x_0)=p(x_1)=0.5$].

On the other hand, suppose one wishes to convey classical
information using a quantum system as a communication channel.
The sender (Alice hereafter) prepares the system in one of
various quantum states
$\rho_i$ with {\em a priori} probabilities $p_i$, so the input signal
is represented by the density matrix $\rho =\sum\limits_i {p_i\rho _i}$.
The intended receiver (Bob hereafter) makes a measurement on the
quantum system, and from its result he tries to infer
which state Alice prepared.
A theorem stated by Gordon \cite{Gordon64} and Levitin \cite{Levitin69}, and
first proved by Holevo \cite{Kholevo73}, asserts that
if Bob is restricted to making
separate measurements on the received states,
then the average information gain is {\em bounded} by
\begin{equation}
I(A:B)\le S(\rho )-\sum\limits_i {p_iS(\rho _i)},
\label{Hol}
\end{equation}
where $S$ is the von Neumann entropy, given by
$S\left( \rho \right) = -\mbox{Tr} \left( \rho \log _2 \rho \right)$.
The equality in (\ref{Hol}) holds if, and only if,
all the transmitted states ${\rho}_i$ commute.
Thus the amount of information
accessible to Bob is limited by the von Neumann entropy of the ensemble
of transmitted states. The maximum von Neumann entropy of an ensemble
of quantum states in a Hilbert space of $n$ dimensions is $n$,
and can be reached only if the ``alphabet'' defined by $\rho$ is a mixture
with identical probabilities of $n$ mutually orthogonal
pure quantum states (called ``letter'' states).
Therefore, as a simple application of the Holevo theorem reveals, the maximum
classical information accessible to Bob when Alice
sends a two-level quantum system (``qubit'') is
one bit. This is what we will refer
to as the {\em Holevo limit}. Achieving the
Holevo limit requires noiseless quantum channels
and perfect detectors, therefore we will assume so hereafter.

Either a classical or quantum $n$-level
system can convey $\log _2n$ bits at the most.
In this sense, quantum communication
is as efficient as classical communication.
However, there is a task that cannot be achieved
by classical means: secure key distribution. Now suppose
Alice wishes to convey a sequence of random classical bits to Bob
while preventing that
any third unauthorized party (Eve hereafter) acquires
information without being detected.
This problem,
known as the {\em key distribution problem}, was first solved by
Bennett and Brassard \cite{BB84} using quantum mechanics.
In recent years many different protocols for
quantum key distribution (QKD) have been
proposed \cite{E91,B92,BBM92,GV95,KI97,CL98,Cabello00b}.
Most of them share the following features:
(i) They need two communication channels between Alice and Bob:
a {\em classical channel} which
is assumed to be public but which cannot be altered.
Its tasks are to allow Alice and Bob to share a code
and information to prevent some kinds of eavesdropping,
to transmit the classical information
required for each step of the protocol, and to
check for possible eavesdropping.
A {\em quantum channel}
(usually an optical fiber or free-space), which must
be a transmission medium that preserves the quantum signals
(usually the phase or the polarization of
photons) by isolating
them from undesirable interactions with the environment. It
is an ``insecure'' channel in the
sense that Eve can manipulate the quantum signals.
(ii) A sequence of $m$ steps. A {\em step} is defined as the minimum part of
the protocol after which one can compute the {\em expected number of secret
bits received by Bob}, $b_s$. Each step consists on an interchange of
a number $q_t$ of qubits (using the quantum channel) and a number $b_t$ of
bits (using the classical channel) between Alice and Bob.
(iii) A test for detecting eavesdropping.
Alice and Bob can detect Eve's intervention by publicly comparing (using
the classical channel) a sufficiently large
random subset of their sequences of bits, which they subsequently discard.
If they find that the tested subset is identical, they can infer that the
remaining untested subset is also identical and secret.
Only when eavesdropping is not found, the transmission is assumed to be
secure.

From the point of view of information theory, a natural definition of {\em
efficiency of a QKD protocol}, ${\cal E}$ is \cite{def}
\begin{equation}
{\cal E}={\frac{{b_s}}{{q_t+b_t}}}\,.
\end{equation}
where $b_s$, $q_t$ and $b_t$ were described above. This definition omits
the
classical information required for establishing the code or preventing and
detecting eavesdropping, because it is assumed to be a constant,
negligible when compared with the number of transmitted secret
bits, $m {b_s}$.
The combination of classical information theory plus the Holevo theorem
imposes an upper limit to the efficiency of any transmission of classical
information (secret or not)
between Alice and
Bob. In particular, they imply that the efficiency of any QKD
protocol is ${\cal E} \le 1$.
The problem addressed in this paper is
whether the limit ${\cal E} = 1$ can be achieved. Or, more generally,
how efficiently random classical information can be distributed between
Alice and Bob (who initially share no information),
while preventing that Eve acquires information without being detected.
As a close inspection of some of the most representative
QKD protocols reveals, so far none of them reaches
the limit ${\cal E}=1$ (see Table I) \cite{comL}.
A QKD protocol with ${\cal E}=1$ requires that Bob can
identify with certainty $n$ different states,
where $n$ is the dimensionality of the Hilbert
space of the quantum channel, ${\cal H}_n$.
Bob can only distinguish $n$ states
with certainty if all of them are mutually orthogonal.
Since there are no $n$ mutually orthogonal mixed states in ${\cal H}_n$,
then the letter states will be necessarily an {\em orthogonal basis of
pure states}.
If the quantum channel is a {\em single} quantum $n$-level system,
the requirements ${b_t}=0$ and ${b_s}={q_t}=\log _2n$ are impossible
to achieve, because then Eve could use the cloning process \cite{WZ82,D82},
to find out what was the state sent by
Alice without being detected.
This problem can be avoided if the quantum channel is
a {\em composed} quantum system.
Then, as was first discovered by Goldenberg and Vaidman \cite{GV95}, the
secret information can be split between the subsystems, so that if Eve has no
access to all the parts at the same time, she cannot recover the
information without being detected. Goldenberg and Vaidman's protocol
was extended and improved by Koashi and Imoto \cite{KI97}.

In this Letter we will present a protocol ${\cal E}=1$
based on
\cite{GV95,KI97} and on the idea of using a larger alphabet that saturates the
capacity of the quantum channel.
Suppose that
the quantum channel is composed by two qubits ($1$ and $2$)
prepared with equal probabilities in one of four orthogonal pure states
$\left\{ {\left| {\psi _i} \right\rangle } \right\}$,
and that
Eve cannot access qubit $2$ while she still holds qubit $1$.
To obtain this ``sequential'' access for Eve,
we can use the configuration in Fig.~1 \cite{GV95,KI97}:
there are two paths between Alice and Bob, one for qubit $1$ and the other
for qubit $2$, and both have the same length $L$.
Alice sends out the two qubits
at the same time.
Qubit $1$ flies to Bob while qubit $2$ is still in a storage ring
(protected against Eve's intervention) of length $l > L/2$.
The aim of this storage ring is to delay qubit $2$ until qubit $1$
has reached the
protected part of the channel near Bob. In that protected part there is
another storage ring of length $l$, so both qubits arrive at the same time
to Bob's analyzer.
To guarantee that Eve has a true sequential access to the two qubits,
Alice and Bob (using the classical channel) must know when qubit $1$
of the first pair will arrive to Bob and
which will be the delay between pairs \cite{KI97}.

Any QKD protocol must fulfill the fact that Eve cannot learn the bits without
disturbing the system in a detectable way. In addition, for practical
purposes, it would be interesting if Bob could easily read
the letter states.
In our protocol, the choice of letter states will be strongly limited by these
requirements. Let us denote as $pnm$ those orthogonal basis of letter states
composed of $p$ product states, $n$ nonmaximally entangled states, and $m$
maximally entangled states. It can be easily seen that
the letter states cannot be a $400$ basis, because then Eve can learn at least
one bit without being detected. For
instance, if the basis was $\left\{ {
\left| {00} \right\rangle,\,
\left| {10} \right\rangle,\,
\left| {+1} \right\rangle,\,
\left| {-1} \right\rangle
} \right\}$, then Eve can learn one bit just by performing a local
measurement on the second qubit and allowing the first one to pass by.
In addition, the letter states cannot be a $004$
basis, because then Eve can learn the two bits without being detected
just by preparing a pair of
ancillary qubits ($3$ and $4$) in a maximally
entangled state, replacing qubit $1$ with qubit $3$, reading the state of the
combined system $1,2$ after receiving qubit $2$, and finally changing the
maximally entangled state of
the combined system $3,4$ by a simple unitary
transformation on particle $4$.
Other possible strategies for eavesdropping in the context of sequential
access, like broadcasting \cite{BCFJS96}, have been investigated by Mor
\cite{Mor98}. Mor's requirement to avoid eavesdropping
(reduced density matrices of the first subsystem must be nonorthogonal and
nonidentical, and reduced density matrices of the second subsystem must be
nonorthogonal \cite{Mor98}) applies to the case when {\em two} (pure or mixed)
letter states are used. As can be easily checked, Mor's condition is
satisfied by at least two pairs of states if one uses an orthogonal basis of
four pure states different than $400$ and $004$. This means that Eve must use
at least two different strategies to obtain information of the key. If for
a particular state she uses
the wrong strategy, Alice and Bob will have a high probability to detect Eve.
Therefore, we conclude that an orthogonal basis of a type different than $400$ or
$004$ can be used as letter states in a QKD protocol with sequential access.
However, these bases present different advantages and disadvantages. On one
side, it will be interesting to use the higher dimension of the quantum
channel to improve the probability of detecting Eve from those protocols using
lower dimensional quantum channels or smaller alphabets. For instance, in a
protocol based on two-letters with the same probability like \cite{BB84}, for
each bit tested by Alice and Bob, the probability of that test revealing Eve
(given that she is present) is $\frac{1}{4}$. Thus, if $N$ bits are tested, the
probability of detecting Eve is $1-\left( \frac{3}{4}\right) ^N$. However,
in a protocol using two qubits as a quantum channel, if Alice and Bob
compare a {\em pair} of bits generated in the same step, the probability for
that test to reveal Eve can be $\frac{3}{4}$.
Thus if $n$ pairs ($N=2n$ bits)
are tested, the probability of Eve's detection is $1-\left( \frac 12\right) ^N$.
However, this improvement is possible only if Eve cannot use
the same strategy to (try to) read two of the four states. This scenario can
be achieved with bases such as $121$, $130$, or $040$.
However, using these bases have a bigger (from an experimental point of view)
disadvantage: as the analysis of some particular cases suggests,
if the qubits are polarized photons, then
Bob cannot discriminate with 100\% success basis like $121$, $130$, or $040$,
using an analyzer with only linear elements (such as beam splitters, phase
shifters, etc.) \cite{John}.
A general proof of this statement for any kind of basis is still an open
problem. Such proof exists for the $004$ bases \cite{VY99,LCS99}.
However, bases such as $202$ or $220$, although they do not improve the
probability of Eve detection, can be used for
QKD in the Holevo limit, and allow Bob to completely discriminate between
the four states {\em without} requiring conditional logical gates, like CNOT
gates between the two qubits, or even electronics to
control conditional measurements on the second qubit depending on the result of
the measurement on the first qubit.
I will present an example of a QKD
protocol in the Holevo limit of this last case.
Consider the following $202$ basis:
\begin{eqnarray}
\left| {\psi_0} \right\rangle & = &
\left| {HH} \right\rangle,
\label{cero} \\
\left| {\psi_1} \right\rangle & = &
{\scriptstyle {1 \over \sqrt{2}}}
\left( {
\left| {HV} \right\rangle +
\left| {VH} \right\rangle
} \right),
\label{uno} \\
\left| {\psi_2} \right\rangle & = &
{\scriptstyle {1 \over \sqrt{2}}}
\left( {
\left| {HV} \right\rangle -
\left| {VH} \right\rangle
} \right),
\label {dos} \\
\left| {\psi_3} \right\rangle & = &
\left| {VV} \right\rangle,
\label{tres}
\end{eqnarray}
where $\left| {H} \right\rangle_i$ means photon $i$ linearly polarized along a
horizontal axis, and $\left| {V} \right\rangle_i$ means photon $i$ linearly
polarized along a vertical axis, and symmetrization is not written explicitly
[for instance, $\left| {HV} \right\rangle$ means
${\scriptstyle {1 \over \sqrt{2}}}
\left( {
\left| {H} \right\rangle_1
\left| {V} \right\rangle_2 +
\left| {V} \right\rangle_1
\left| {H} \right\rangle_2
} \right)$].
Alice prepares one of the four states (\ref{cero})-(\ref{tres}) and
sends them out to Bob using a setup to guarantee Eve's sequential
access (Fig.~1). The two qubits arrive at Bob's state analyzer at the same
time. In the case of the four photon polarization states
(\ref{cero})-({\ref{tres}), Bob analyzer to discriminate with 100\%
(theoretical) success between the four states can be
realized in a laboratory using a $50/50$ beam
splitter, followed by two polarization beam splitters (which transmit
horizontal polarized photons and reflect vertical polarized photons), and four
detectors \cite{dc2,telep2}; see Fig.~2. After the polarization beam splitters,
as a simple calculation (up to irrelevant phases) reveals, the four states
(\ref{cero})-(\ref{tres}) have evolved into
\begin{eqnarray}
\left| {\psi_0} \right\rangle & \rightarrow &
{\scriptstyle {1 \over \sqrt{2}}}
\left( {
\left| {D_{1} D_{1}} \right\rangle -
\left| {D_{3} D_{3}} \right\rangle
} \right),
\label{fincero} \\
\left| {\psi_1} \right\rangle & \rightarrow &
{\scriptstyle {1 \over \sqrt{2}}}
\left( {
\left| {D_{1} D_{2}} \right\rangle -
\left| {D_{3} D_{4}} \right\rangle
} \right),
\label{finuno} \\
\left| {\psi_2} \right\rangle & \rightarrow &
{\scriptstyle {1 \over \sqrt{2}}}
\left( {
\left| {D_{2} D_{3}} \right\rangle -
\left| {D_{1} D_{4}} \right\rangle
} \right),
\label {findos} \\
\left| {\psi_3} \right\rangle & \rightarrow &
{\scriptstyle {1 \over \sqrt{2}}}
\left( {
\left| {D_{2} D_{2}} \right\rangle -
\left| {D_{4} D_{4}} \right\rangle
} \right),
\label{fintres}
\end{eqnarray}
where $\left| {D_{i} D_{j}} \right\rangle$ means one
photon in detector $D_i$ and the other in detector $D_j$,
again symmetrization is not written explicitly [for instance,
$\left| {D_{1} D_{2}} \right\rangle$ means
${\scriptstyle {1 \over \sqrt{2}}}
\left( {
\left| {D_1} \right\rangle_1
\left| {D_2} \right\rangle_2 +
\left| {D_2} \right\rangle_1
\left| {D_1} \right\rangle_2
} \right)$]. Thus, a single click on detectors $D_1$ or $D_3$ ($D_2$
or $D_4$) signifies detection of $\left| {\psi_0} \right\rangle$
($\left| {\psi_3} \right\rangle$), while two clicks, one on $D_1$ and
the other on $D_2$, or one on $D_3$ and the other on $D_4$
(one on $D_2$ and the other on $D_3$, or one on $D_1$ and the other on $D_4$)
signifies detection of $\left| {\psi_1} \right\rangle$
($\left| {\psi_2} \right\rangle$).

Long storage rings have a low efficiency for the transmission of polarized
photons, so other methods to achieve sequential access must be developed in
order to perform QKD in the Holevo limit for long distances.
For instance, Weinfurter
has suggested \cite{Harald} using momentum-time entangled photons and a
Franson-type device \cite{Franson}.
On the other hand, QKD protocols with ${\cal E}=1$ can be extended
to quantum channels composed of $n \ge 2$ subsystems with $m \ge 2$ levels,
supposing Eve has only a sequential access to the subsystems.
If $nm \ge 6$
Alice could use even a basis with only product states \cite{BDF99}, although
then Bob would need some quantum interaction between the subsystems
in order to achieve a complete discrimination of the letter states
\cite{palma}.

The author thanks J. Calsamiglia,
G. Garc\'{\i}a de Polavieja, H.-K. Lo, and H.
Weinfurter for very useful comments and suggestions.
Part of this work has been done at Cambridge University
and at the Sixth Benasque Center for Physics under
support of the University of Seville
(Grant No.~OGICYT-191-97)
and the Junta de Andaluc\'{\i}a.
(Grant No.~FQM-239).

\begin{table}[tbp]
\begin{center}
\begin{tabular}{lcccc}
\hline
\hline
\multicolumn{1}{c}{Scheme} & $b_s$ & $q_t$ & $b_t$ & ${\cal E}$ \\ \hline
Bennett, 1992 \cite{B92} & $< 0.5$ & $1$ & $1$ & $< 0.25$ \\
Bennett and Brassard, 1984 \cite{BB84} & $0.5$ & $1$ & $1$ & $0.25$ \\
Goldenberg and Vaidman, 1995 \cite{GV95} & $1$ & $2$ & $\ge 1$ & $\le 0.33$
\\
Ekert, 1991 \cite{E91,BBM92} & $1$ & $1$ & $1$ & $0.5$ \\
Koashi and Imoto, 1997 \cite{KI97} & $1$ & $2$ & $0$ & $0.5$ \\
Cabello, 2000 \cite{Cabello00b} & $2$ & $2$ & $1$ & $0.67$ \\
\hline
\hline
\end{tabular}
\end{center}
\noindent TABLE I. {\small Efficiency $\cal E$
of different QKD protocols.
In Goldenberg and Vaidman's protocol, $b_t$ contains
the sending time of the qubits.
In Ekert's protocol the values refer to a
more efficient version, suggested by Ekert to the authors of Ref.~\cite{BBM92},
in which Alice tells Bob her choice before Bob's measurement.
Both in Goldenberg and Vaidman's, and in Koashi and Imoto's protocols, $q_t$ is
taken to be $2$ because their quantum channel is a photonic state in {\em two}
paths, which is a four-dimensional Hilbert space, although their letter states
do not span the whole Hilbert space.}
\end{table}


\begin{figure}
\epsfxsize=8.6cm
\epsfbox{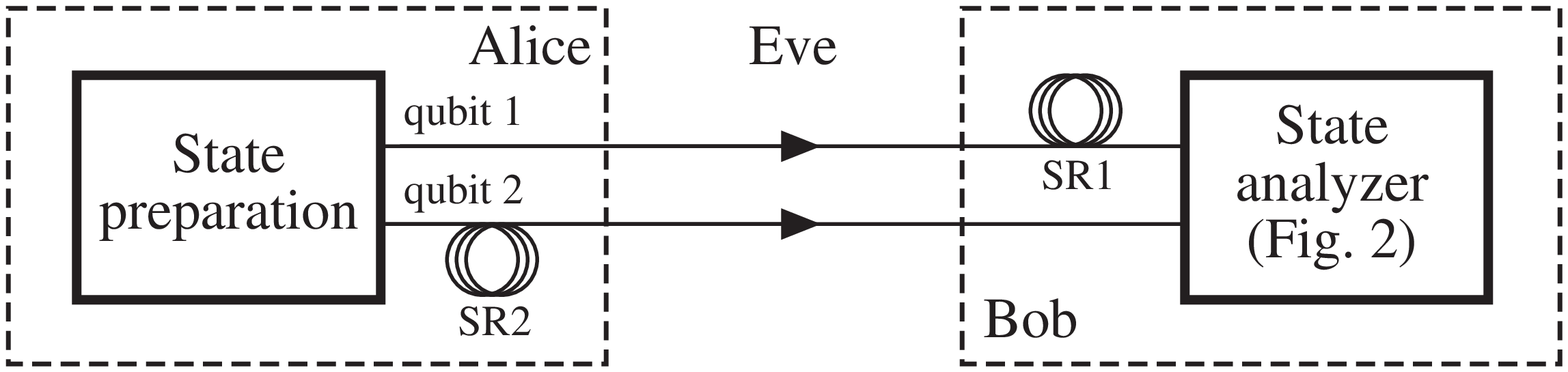}
\end{figure}
\noindent FIG.~1: {\small Scheme to force that Eve has only a
sequential access to the two qubits.}
\begin{figure}
\hspace{0.2cm}
\epsfxsize=7.2cm
\epsfbox{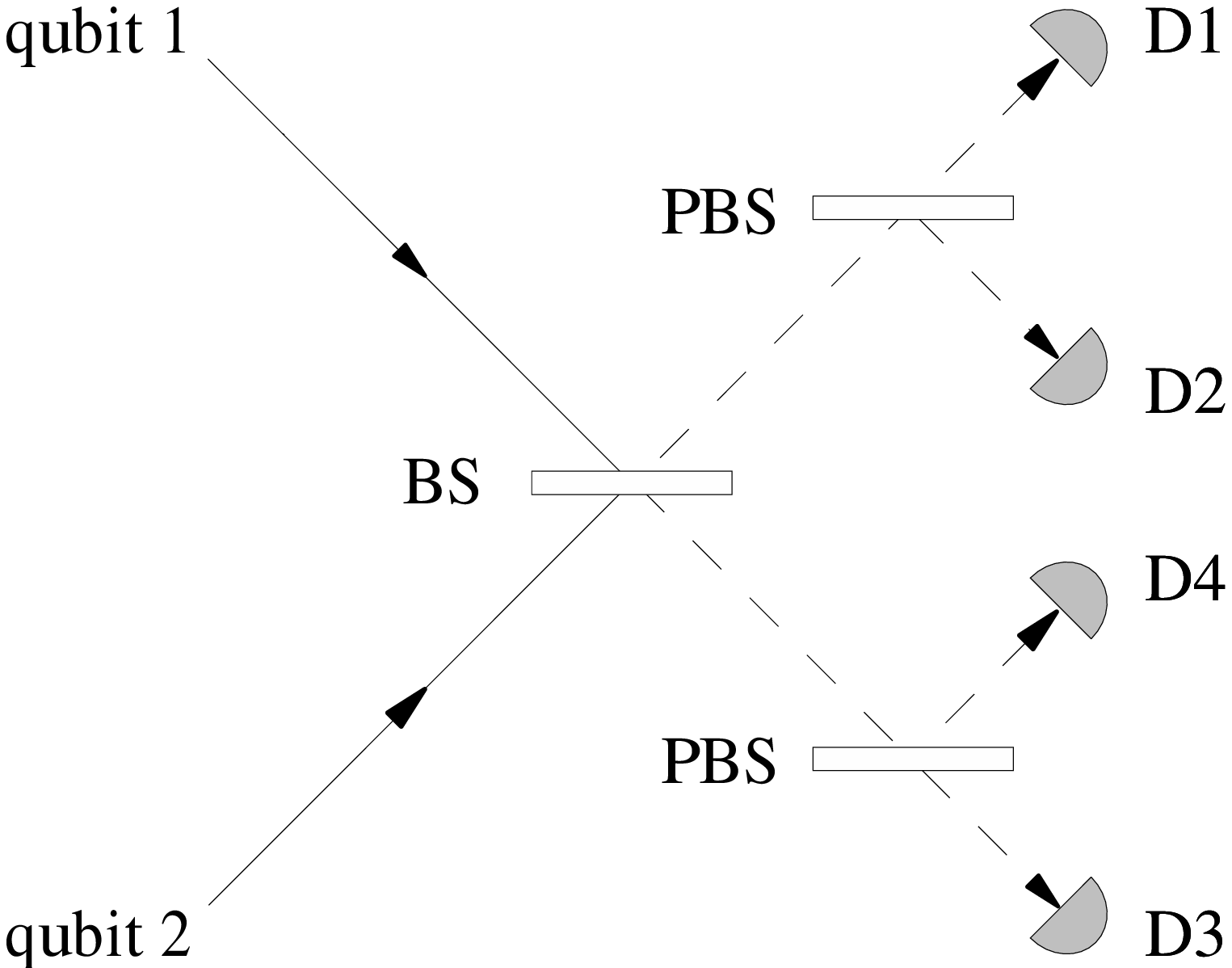}
\end{figure}
\noindent FIG.~2: {\small Scheme of Bob's analyzer to discriminate
unambiguously between the four states (\ref{cero})-(\ref{tres}).}


\begin{references}
\bibitem{def} In this definition it is assumed
that the cost of transmitting qubits is the same as
the cost of transmitting
bits, which is not realistic in practice.
\end{references}
\end{document}